# Making the road by searching – A search engine based on Swarm Information Foraging

Daniel Gayo-Avello[1]

Department of Computer Science, University of Oviedo, Edificio de Ciencias, C/Calvo Sotelo s/n, 33007 Oviedo, Spain

David J. Brenes

Indigo Group, C/Campoamor 28 1º Oficina 5, 33001 Oviedo, Spain

## Abstract
Search engines are nowadays one of the most important entry points for Internet users and a central tool to solve most of their information needs. Still, there exist a substantial amount of users' searches which obtain unsatisfactory results. Needless to say, several lines of research aim to increase the relevancy of the results users retrieve. In this paper the authors frame this problem within the much broader (and older) one of information overload. They argue that users' dissatisfaction with search engines is a currently common manifestation of such a problem, and propose a different angle from which to tackle with it. As it will be discussed, their approach shares goals with a current hot research topic (namely, learning to rank for information retrieval) but, unlike the techniques commonly applied in that field, their technique cannot be exactly considered machine learning and, additionally, it can be used to change the search engine's response in real-time, driven by the users behavior. Their proposal adapts concepts from Swarm Intelligence (in particular, Ant Algorithms) from an Information Foraging point of view. It will be shown that the technique is not only feasible, but also an elegant solution to the stated problem; what's more, it achieves promising results, both increasing the performance of a major search engine for informational queries, and substantially reducing the time users require to answer complex information needs.

## 1. Introduction
Web searching is becoming a common habit among Internet users: in fact, almost half (49%) of the users in the United States employ search engines on a typical day [74]. However, the users' degree of satisfaction with major Web search engines is, to the best of our knowledge, largely unsettled, and we must rely in rather small and/or close-world studies.

Thus, according to Jansen *et al.* [44] and Silverstein *et al.* [89] most users solve their information needs with two queries or less (67% of the users according to the first study and 78% according to the second one); they do not usually go beyond the first page of results (58% and 85% of the users, respectively); and, according to Jansen and Spink [43] 66% of the users review 5 results or less, and 30% of the users review just one result. Thus, Jansen and Spink argue that (1) most users' information needs are not complex, (2) the first results provided by the search engines are actually relevant to the original information needs, and (3) on average 50% of the results are relevant to the users.

---
[1] Corresponding author. E-mail address: dani@uniovi.es

Therefore, the subtext in that study is as follows: on average 50% of the results provided by the search engines are not relevant and, in addition to that, there exist complex information needs that, in all probability, are obtaining a larger number of irrelevant results. This is consistent with a small evaluation conducted by Hawking *et al.* [40] where all of the twenty search engines evaluated returned, on average, less than 60% of relevant results.

There are additional indirect evidences about the users' degree of satisfaction with search engines. For instance, Fox *et al.* [34] describe a machine learning approach to infer such a degree from user actions (e.g. time spent on a page, use of scrolling, number of visits to a certain page, etc.) In order to train their system they collected data from actual users including explicitly stated feedback about the search engine results. According to their report, 28% of the search sessions were reported by the users as unsatisfactory and 30% only partially satisfactory.

Another recent study performed by Xu and Mease [101] reveals that, on average, it takes users about 3 minutes to finish a search session, and they are mostly dissatisfied when the search session lasts longer than that time. Besides, it must be noticed that in that study users were instructed to finish their sessions not only when they found an answer but also when they thought a typical user would give up; thus, the 3 minute time span includes abandonment without achieving the goal of the information need.

So, to sum up, web searches are becoming the starting point for many Web-mediated activities [52] and many of them reveal a mainly informational goal [15, 84]; that is, the users resort to search engines to find factual data on the Web. However, the users' search experience is far from being perfect and, indeed, a substantial amount of searches should be considered unsatisfactory and, even, failed searches.

Hence, it is not surprising that considerable effort it being devoted to improve the relevancy of the results provided by search engines. It is also unsurprising that most of such efforts rely on the exploitation of the search engine query-click logs (files storing the users' interactions with the search engine; a fragment from such a log is shown in Figure 1). Joachims [47] was the first one to employ such data as implicit relevance judgments about the results retrieved by the search engine. He used that data to learn the ranking function of a specialized meta-search engine which later outperformed a major search engine. Following that seminal work, the exploitation of query-click logs to improve search engine ranking functions has become a popular matter of research[2]. We will return on that in the Literature Review section.

Thus, this work addresses the problem of transparently improving the users' Web search experience by exploiting the previous actions of other individuals. In this sense, our proposal shares the same mission with the learning-to-rank community but it addresses that goal from a different perspective. As it will be shown, our technique applies swarm intelligence concepts, concretely from ant algorithms, from an information foraging point of view. It is worth noting that our method departs from most of the work in the field of swarm intelligence because we are not using artificial agents but actual human beings.

---

[2] Up to now, three workshops on the topic of learning to rank for information retrieval have been held within the ACM's SIGIR conferences.

The paper is organized as follows. First of all, a comprehensive literature review regarding the information overload problem, in general, and Web searching, in particular, is provided. Then, the research questions driving the study are posed, and a new swarm-based, adaptable, meta-search technique is proposed and described. After that, the experimental framework in which this study was conducted is described: namely, an offline evaluation and an online evaluation by means of a controlled experiment. Afterwards, results from both evaluations are provided and discussed, along with the implications and limitation of this study.

## 2. Literature review

Although stated in terms of user dissatisfaction with search engines performance, the root of the problem we are tackling with in this paper is information overload. Definitely, it is not a new one but the advent of the Web in the early 1990s did certainly amplify it and made it public knowledge. In fact, search engines are just one of the tools devised to help users to solve their information needs in the Web and, as we have noted, there is still much room for improvement. For this reason, we feel that, in order to paint the big picture, this literature review must cover, although briefly, some previous approaches even when they are only marginally related to the specific motivation of our work. Hence, this section addresses the following points: the Web as an Information Retrieval system; software agents and collaborative filtering; information foraging on the Web; learning to rank for information retrieval and; finally, other approaches to build adaptable Web search systems.

### 2.1. The Web as an Information Retrieval system

The first web server (`info.cern.ch`, formerly `nxoc01.cern.ch`) came into operation in 1990 and by the end of 1992 there were about 20 web servers [11]. At that moment it was relatively easy to manually keep a website directory and, in fact, the CERN and the NCSA were managing such indices. However, the Web was rapidly expanding [39]: by the end of 1994 there were more than 10,000 web servers and, therefore, manually maintaining a Web directory was increasingly cumbersome.

The lack of a proper way to find servers and documents in the early Web was already a problem and different systems were being developed to tackle with it. Of particular note are ALIWEB [50] and WWW Worm [61]. Both systems relied on the now common idea of automatically crawling the Web to create a document database which, later, would be queried by the users. Other representative contemporary systems were Jumpstation [33], Wanderer [39], WebCrawler [75], and Lycos [60].

Nonetheless to say, the results provided by these primitive search engines were, for today's standards, rather poor. It would be Jon Kleinberg [49] who laid the foundations of hyperlink analysis which is the base for modern Web search engines such as *Google* [14].

In the last decade Web search engines have become central for most Web-mediated activities; however, there were other techniques and tools which were devised to help users to solve their information needs on the Web.

### 2.2. Software agents and collaborative filtering

Software agents were a heavily investigated topic in the 1990s in relation to the information overload problem. A few of the most representative works in the area are those by Patti Maes [59], Morita and Shinoda [67], Menczer *et al.* [63], Henry Lieberman [55], Balabanobic *et al.*

[8], Balabanovic and Shoam [6, 7], Pazzani *et al.* [72, 73], Goldman *et al.* [38], or Mouchas and Zacharia [68].

In all these works software agents were intended to filter or recommend content to the users (e.g. e-mail, USENET posts, books, music, or web pages). All of them relied on a user profile (usually a weighted term vector) that could be obtained either explicitly (asking the user to rate some previous contents or the filtered/recommended documents [6, 7, 8, 59, 63, 68, 72, 73]), or implicitly (observing the user actions: clicks, scrolling, etc. [55, 67]). Then, the user profile was employed by the agent to filter content [6, 38, 59, 63, 67, 73] or to find new relevant content on its own [7, 8, 55, 68, 72].

These approaches faced the information overload problem mainly from a personalization point of view, without exploiting the collective behavior (except for the work by Balabanovic and Shoam [7]). In addition to that, although many of them rely on Web search engines their goal was not to improve the search engine performance but filter the provided results according to the users' individual preferences.

The first system providing collaborative filtering was the Tapestry mail system at Xerox PARC described by Goldberg *et al.* [37]. In their own words: *"collaborative filtering simply means that people collaborate to help one another perform filtering by recording their reactions to documents they read."*

Thus, collaborative filtering can be considered orthogonal to content based filtering which was used by all of the systems previously described. Following the work by Goldberg *et al.* several systems applied collaborative filtering to fight information overload –e.g. GroupLens [82] or Ringo [88]. However, most of such systems still require users to explicitly provide ratings or feedback about the recommended resources.

A much simpler (and smarter) approach was devised by Rucker and Polanco [85] in their Siteseer system. Siteseer collected users' bookmarks and used them both to find similar users, and to recommend such bookmarks to other users which were unaware of them. The PHOAKS system [93] also exploited available data provided by the users, in this case in USENET. PHOAKS mined USENET groups in order to find posts containing URLs and recommendations; then, such URLs were ranked according to the number of posts recommending them.

### 2.3. Information foraging on the Web

Thus, by the end of the 1990s a few methods [85, 93] had been proposed to exploit the users' collective behavior in order to suggest relevant resources to those users and, which is equally important, without requiring the users to perform any extra action. However, although in both cases the system was seeking Web resources, it was not exploiting the Web users behavior (Siteseer employed browser bookmarks and PHOAKS mined USENET posts).

At this point, it is worth noting that several theories have tried to describe the users' information seeking behavior in complex information systems such as the Web: ASK (Anomalous State of Knowledge) [10], Berrypicking [9], and Information Foraging [76, 77] are among the most relevant ones. The later is heavily inspired by the optimal foraging theory; in fact, one of its key concepts is the so-called "information scent": the user's perception of the cost, value, and access path to a distal resource obtained by proximal cues (e.g. links, snippets, tags, comments, etc.). The application of Information Foraging ideas to the Web seems quite natural [22] and several approaches to improve Web information seeking and retrieval can be traced back to this theory. Hence, we will briefly describe a few pertinent works in such a line.

For instance, the MEMOIR system [26] mined the trails left by each user (i.e. the actions on documents they visited) in order to detect other similar trails. Such information allowed users to find colleagues with related interests in addition to new relevant documents. Aggarwal [1] described a system to discover topically specific resources by mining the traces left by Web users in HTTP proxies.

However, perhaps the most explored line of research has been augmenting the information provided by hyperlinks in order to suggest the users the most promising paths to achieve their goals. Thus, WebWatcher [4] allowed users to specify some keywords and, then, it recommended the most promising links within the current web page according to previous user sessions. Golden Path Analyzer [3] clustered click-streams within a website to find the shortest successful path to solve a task, and then suggested such golden-paths to users whose clicks matched a given cluster. ScentTrails [69] allowed users to introduce keywords at any moment and, then, it enriched the links providing a path to achieve the goals described by those keywords. Finally, the method by Wu and Aberer [98] also worked on a single website augmenting the information provided by hyperlinks with a method inspired in the ant foraging behavior (i.e. heavily clicked links were recommended in favor of less visited links).

### 2.4. Learning to rank for information retrieval

Learning to rank is at this moment a hot topic of research; the goal of which is to automatically learn ranking functions from training data, usually in the form of click-through data. Joachims outlined in a seminal paper [47] the key aspects of this research topic: (1) explicit feedback from users cannot be taken for granted, which is more, it is largely unnecessary because enough information is already available in the query logs from the search engine; (2) in fact, users' clicks on the results can be employed as relevance judgments albeit not in an absolute scale; and (3) a machine learning method (in his case SVM) can be trained on such data to obtain a new ranking function which can outperform state-of-the-art search engines.

Apart of SVMs many other machine learning methods have been adapted for the topic. This is a certainly non-exhaustive list of the devised algorithms up to date: RankBoost [35], RSCF [92], RankNet [18], QBRank [105], GBRank [104], AdaRank [100], FRank [94], LambdaRank [17], ListNet [19], and McRank [54].

There exist, however, two main drawbacks with learning to rank approaches. First, the training phase is computationally expensive and, thus, fast methods are being sought (e.g. [31]). Second, it mostly applies offline learning methods, that is, when more training data is available the system must be trained again. Because such training data is not other thing than click-through data this situation is commonplace and, thus, online learning methods are also investigated (e.g. [20, 80, 81]).

On a side note, not only machine learning methods have been used to learn ranking functions but also soft computing techniques such as genetic programming [102] and, also, swarm intelligence [27].

### 2.5. Other approaches for adaptable Web search engines

Up to now we have described both systems in which users' behavior is mined to personalize their searching/browsing experience (e.g. by filtering or recommending resources, or by giving cues on the links to follow in a website), and methods to learn new ranking functions for search engines based on the click-through data. Nevertheless, there exist several other relevant proposals to achieve ranking mechanisms based on the users' behavior[3] which do not fit within the aforementioned research lines.

Probably the first Web search engine to incorporate usage behavior into the ranking function was the, now defunct, DirectHit [25]. However, the details of its implementation are not well known and, thus, we could equally well start with the work by Ding and Chi [28]. In order to exploit the users' behavior they devised an ad hoc ranking formula with three main components: (1) content-based ranking, (2) link-based ranking, and (3) usage-based ranking. This last one took into account every click users did on a result in response to a specific query in addition to the individual starting access times and the time spent on that result.

Another well-known method to use clicks for ranking is MASEL [103] which exploits the mutual reinforcement between users, queries and resources. Simply stated, good (i.e. experienced) users issue good queries which, in turn, reach good (i.e. relevant) documents which are then clicked by good users. MASEL mines the click-through data in order to detect those good users, queries and, finally, documents.

Baeza-Yates *et al.* [5] proposed a ranking method in which the relevance of the results is boosted according to the preferences exhibited by previous users. Their method relied on a previous clustering phase where similar queries are grouped; then, URLs are extracted from these clusters, and ranked according to the users' clicks. Another system with many resemblances to this work is I-SPY [90].

The approach by Agichtein and Zheng [2] is totally different; instead of devising a method to exploit usage behavior to boost results they used that behavior to detect potentially top-list results. Their work is highly related to that by Piwowarski and Zaragoza [78], and White *et al.* [97].

Yet another different method was developed by Craswell and Szummer [23]. They suggested applying the Markov random walk model (the base of the PageRank algorithm [14]) to a graph obtained from the click-through data. Such a click graph uses queries and documents as nodes, while edges represent clicks linking a given query to a given document. This way, their method is not only able to find the most relevant documents for a given query but also the opposite, the most relevant queries (i.e. keywords) for a given document.

BrowseRank [57] is slightly related to this work because it also applies a Markov model to a graph derived from click-through data. However, unlike Craswell and Szumer, Liu *et al.* just take into consideration the documents and, thus, they are actually exploiting the browsing behavior but not the users' searching behavior.

To conclude this section, it must be noticed that none of the aforementioned techniques (except perhaps [28, 90]) can be applied to adapt the search engine to users in real time. In fact, all of them should be periodically rerun to modify the search engine's response according to the last recorded usage behavior.

---

[3] In this section we will cover those methods that cannot be properly considered online machine learning to rank.

# 3. Research motivation

## 3.1. Research questions

We have stated that information overload is still a common problem for Internet users. Hence, we have described several approaches to help users to fight such a problem: software agents, implicit feedback, collaborative filtering, and assisted browsing and searching.

In fact, because of the pervasiveness of Web search today, we have focused in the commonest current materialization of information overload: the substantial amount of user searches unsatisfactory solved by search engines. With regards to this situation, we have then reviewed several methods to improve search engines' performance, such as learning to rank and different methods to mine the users' search behavior.

In relation to this, we have pointed out several key aspects of the matter: First, the foraging nature of the users' information seeking behavior. Second, the need for search engines to adapt to such a behavior in order to modify their response. And, third, the need of performing such an adaptation in real time.

Almost none of the methods aforementioned in the Literature Review cover all those three aspects –in particular real time adaptation; and those which perhaps could (i.e. [28, 90]) would greatly benefit from a more thorough application of Information Foraging ideas.

We felt that a swarm-based approach –similar in spirit to that of Wu and Aberer [98] or Olston and Chi [69]– could indeed satisfy all those three requirements while being, at the same time, far more comprehensible and elegant than *'ad hoc'* methods.

Thus, the main research questions addressed in this study are the following: (1) Is it feasible to implement a swarm-based real-time adaptable search engine? (2) For which kind of queries, if any, does the search engine's performance increase? And, (3) does the user search experience improve in any measurable way?

The first research question was directly addressed by these authors by proposing a new technique that will be described in the following subsection. The evaluation of such a technique, later described, will provide answers for the second and third research questions and, thus, will completely answer the first one.

## 3.2. A model for a swarm-based real-time adaptable search engine

Simply stated, Swarm Intelligence [12] refers to the emergence of "intelligent" behaviors from groups of simple and/or loosely organized agents. Ants represent an archetypical example of swarm intelligence and their use of pheromone trails inspired the well-known family of ant algorithms [29].

As we have previously stated, we are interested in the adaptation of ant colony search strategies to build adaptable search engines. Certainly, we the users are more intelligent than ants but the size of the Web overwhelms us as individuals when we browse it. Yet, complex swarm behaviors actually arise and, in fact, collective works such as the *Wikipedia*, *del.icio.us*, or the Web itself are stigmergic processes. So, in short, the users are collectively exploring the Web and issuing relevance judgments every time they submit a query and click a result on a search engine. Thus, a swarm-based approach seems a suitable idea to make such systems able to exploit the seeking behavior of the users.

We will now very briefly describe the ants' pheromone trails because they constitute the basis of our proposal. Each ant leaves a trail of pheromones which can be identified and followed by other ants of its anthill. These trails are not permanent but evaporate at a constant pace. When foraging for food, individual ants browse their environment randomly at first and later following the strongest pheromone trails. Since these trails evaporate with time, the closer the source of food the stronger the trail and, this way, the ant swarm eventually finds the shortest path from the anthill to the food.

It is easy to find parallelisms between the way an anthill forages for food and the way users forage for information by using search engines; with the obvious exception of the lack of pheromone trails in the case of Web users.

Without pheromone trails, search engine users are forced to browse through snippets, and click results until they eventually find what they were looking for, no matter other users had found it before them. This would be irrelevant if such a scenario was infrequent but, indeed, it is just the opposite: about 30% to 40% of queries have been previously submitted, and the most popular ones are shared across many different users [99].

Now let's suppose that the search engine allowed users to follow "pheromone trails" left by previous individuals. Clearly, heavily visited results should be reinforced and, thus, some kind of link enrichment could be provided. However, doing that in a way similar to [3, 4, 69, 98] would be a wrong approach in all probability. Let's see why: imagine that the most promising result for a given query appears beyond the first page of results; no matter the signals the search engine uses to highlight that link, most of the users will miss it because they just don't go beyond the first page of results.

Hence, those heavily visited links should be boosted to appear in the first page of results. This could seem superficially similar to the methods described in [2, 5, 78, 90, 97]. However, the goal of such methods is to detect top-list results for given queries; unlike them, we feel that a stochastic approach[4] could be better, besides perfectly fitting within a swarm-based method. In other words, heavily visited results are more likely to appear in the first places but less visited results still can appear and, thus, they have a chance to be clicked and, thus, reinforced.

Now, we will provide a short formalization of our model. In first place, we assume that users' interactions with the search engine are available under the form of query sessions[5] in the sense of Wen and Zhang [96]:

> *"A query session is made up of the query and the subsequent activities the user performed."*

In other words, a session consists of one query and several (or none) clicked documents:

$$S := \left(Q, (D)^*\right)$$

---

[4] At this moment the probability of recommending a document *d* for a specific query *q* is directly related to the pheromone weight for *(q,d)*; in the future we would try to adapt the edge selection strategy commonly used in ant colony optimization to this recommendation problem.

[5] Other definitions of search session exist; however, a discussion on the differences between them, and the methods to sessionize a query log is out of the scope of this paper. We suggest the interested reader to refer to the work by Gayo-Avello [36].

Thus, given the sets of available documents $D$ and known queries $Q$ the pheromones $\varphi$ associated with the pairs of documents and queries are defined as:

$$\varphi : (Q, D) \rightarrow W \times T$$

Where both $W$ and $T$ are non-negative values representing, respectively, the weight or intensity of the pheromones for a given query-document pair –from now on $\varphi_W(q,d)$– and the last time a specific document was clicked as a result for a given query –from now on $\varphi_T(q,d)$.

It is worth remembering that we are not using artificial ants but real users as agents; thus, we depart at this point from the common definition of trail intensity and, instead, we separately define pheromone deposition and evaporation. In our proposal every time a user clicks a document $d$ as a result for a query $q$ or, using a swarm analogy, every time a user traverses the path $q \rightarrow d$, s/he deposits some pheromone on that path. At this moment, we are using the unity for every user every time but it could be interesting to analyze other options:

$$\varphi_W(q,d) := \varphi_W(q,d) + 1$$

Pheromone evaporation is obtained by means of exponential decay ($t$ is the current time):

$$\varphi_W(q,d) := \varphi_W(q,d) \cdot e^{\lambda(t - \varphi_T(q,d))}$$

The previous equation can be transformed into a far more intuitive half-life version by assuming:

$$\lambda = \frac{\ln\left(\frac{1}{2}\right)}{\delta}$$

This way, we have the $\delta$ parameter which is the time required for the pheromone weight to fall to one half of its initial weight. For the experiments described in this work $\delta$ values of 86,400s (one day) and 604,800s (one week) have been employed.

$$\varphi_W(q,d) := \varphi_W(q,d) \cdot \left(\frac{1}{2}\right)^{\frac{t - \varphi_T(q,d)}{\delta}}$$

Pheromone evaporation can be straightforwardly implemented by decaying every pair *(q,d)* periodically but more efficient implementations are possible. Nevertheless, evaporation must always be computed before deposition.

By applying this simple model it is possible to have a weighted ranking of documents associated with queries. In spite of the resemblance with the methods described in [2, 5, 78, 90, 97] our proposal has two key differences: unlike those methods, it can be applied in real-time and, more importantly, it incorporates the decay of users' attention[6] in a completely natural way thanks to the evaporation mechanism.

---

[6] To the best of our knowledge the only work which incorporates the need to "forget" in order to adapt to the constantly changing users' interests is the one by Koychev and Schwab [51].

This model was implemented with slightly variations in three different flavors: *Naïve*, *Ranking-bias* and *Elaborate*. The *Naïve* version corresponds to a straightforward implementation of the model; the so-called *Ranking-bias* and *Elaborate*, on the other hand, incorporated different ideas from the literature aiming to better reflect some observed user behaviors.

In the *Naïve* version each time a document is clicked, the pheromones increase in a fixed amount, no matter the position of the document within the page of results. That is, if the user clicks in the first result the pheromones for that document increase in one unity and the same occurs if the user clicks in the last document of the third page of results. This, however, is not fair because of the so-called ranking bias: the fact that users tend to click much more in the highly ranked results even if they are less relevant than other documents [48].

Hence, we explored several models which aimed to explain such ranking bias, namely, the User Persistence Model [70], the Single Browsing Model [30], and the Cascade Model [24]. Suffice it to say that none of them perfectly fits the behavior observed in actual query logs and, hence, for this initial work we eventually decided to base on the Single Browsing Model by Dupret and Piwowarski.

Simply stated, such model describes a user who goes down the search results sequentially, for each snippet s/he can or cannot decide to read it (with a certain probability) and, in case of reading it, s/he can or cannot decide to click on it (depending on the attractiveness of the snippet). Besides, the probability of examining a result on a given position depends on the position of the last result the user has clicked on. Dupret and Piwowarski computed those probabilities for a major search engine's query log and, hence, it was pretty easy to simulate such browsing behavior using that data.

So, in short, the *Ranking-bias* version does not increase pheromones by one unity on every click but according to the aforementioned examination probabilities. For instance, according to Dupret and Piwowarski's data, a user who does not click either the first, or the second, or the third result will click the fourth one with probability 0.82. Thus, in that case the *Ranking-bias* method would increase the pheromones for that document not in 1 but in 1.22 ($0.82^{-1}$).

The *Elaborate* version, on the other hand, was partially inspired by the work by Joachims *et al.* [48]. They described several strategies to derive pairwise preferences from clicks. One of them is the so-called *"Click > Skip above"* which, simply stated, assumes that when a user clicks on a document after skipping several others, it is because s/he prefers that document over the skipped ones. For instance, let's suppose a user who clicks on documents 1, 3, and 5 from a list of 10: $d_1^*, d_2, d_3^*, d_4, d_5^*, d_6, d_7, d_8, d_9, d_{10}$. According to such a strategy: $d_3>d_2, d_5>d_2, d_5>d_4$, and none other preferences can be derived.

In our implementation we made two extra assumptions: (1) the user is satisfied by the ordering of the clicked results, and (2) any document below the last clicked result is not examined by the user. Thus, from a click-list such as $d_1^*, d_2, d_3^*, d_4, d_5^*, d_6, d_7, d_8, d_9, d_{10}$ we derive the following "ideal" ordered list: $d_1^*, d_3^*, d_5^*, d_2, d_4$. Then, the pheromones are updated but, unlike the *Naïve* and *Rank-bias* versions, their amounts are not only associated with each query and document but also with the assumed user preferred ranking. Thus, in the current example we would increase the pheromone amount for $d_1$ at position 1, $d_3$ at position 2, $d_5$ at position 3, $d_2$ at position 4, and $d_4$ at position 5. We will later discuss the results achieved by each of these different flavors; however, it must be said that the idea of demoting the rank of skipped documents did not seem to affect performance in any significant way and, hence, we eventually limited ourselves to clicked results[7].

So, to sum up, our proposal for a swarm-based adaptable search engine relies on pheromone trails stored for every query-document pair[8]. Those pheromones increase by a fixed amount[9] every time a user clicks on a document as a result of a query; and, in addition to that, pheromones evaporate with time, thus, providing a natural way to incorporate attention decay and interests shifts. Then, when a user submits a known query, one or more results can be recommended from those previously clicked. Such a recommendation depends on the amount of pheromones deposited in the document but it is performed in a stochastic way to still give not heavily visited results a chance to be examined by the users.

## 4. Research design

The main goal of this study was to check for the feasibility of swarm-based real-time adaptable search engines. To attain that, we implemented several prototypes for such a system based on the different flavors of the model described in the previous section. Then, in order to address the second and third research questions, regarding the performance of that system and the eventual impact in user experience, we conducted offline and online experiments. The former consisted of an evaluation of the different flavors of the technique on a real query log; the later was a controlled experiment involving 20 volunteers. Consequently, this section describes the dataset employed in the offline experiment, details the performance measures to evaluate the system, and, finally, describes the setup of the controlled experiment.

### 4.1. Offline evaluation against the AOL query log

In an ideal situation an information retrieval system (e.g. a search engine) should only provide relevant results for the submitted queries. However, in the practice we accept that the goal for such a system is to find the largest amount of relevant documents with the smallest possible number of irrelevant ones. Relevance is a subjective matter but, under experimental conditions, it is not a problem because it is always possible for a panel of experts to devise a test document collection, in addition to sets of test queries and relevant documents for each query. Thus, suffice it to say that there exists a long tradition of evaluation in information retrieval by means of test collections comprising documents, queries, and relevance judgments (see [83]).

Of course, the development of such collections requires a great amount of work. Besides, it would be extremely difficult to develop such a kind of collection (i.e. including queries and judgments) to reflect real usage of search engines. Hence, some recent research (e.g. [46, 56]) has relied on the use of click-through data to evaluate search engine performance. The basic idea is indeed quite simple: to take clicks as positive relevance judgments, i.e. if a user clicks on a result after submitting a query s/he is considering that document somewhat relevant for the query.

Therefore, for the offline evaluation of our proposal we relied on the AOL query log [71]. This dataset, released by AOL on August 2006, contains more than 30 million queries sampled over three months from over 650,000 users and, to the date, it is the largest and most complete query log publicly available to scholars.

---

[7] This is coherent with the findings of Schwab et al. [87] who showed that systems using only positive evidence are comparable to those using both positive and negative evidence

[8] Or every query-document-position triplet in the case of the *Elaborate* method.

[9] Except for the *Rank-bias* method in which the increase depends on the position of the document within the result list and the position of the last clicked result.

### 4.1.1. Data description

Needless to say, we did not use the whole AOL dataset to evaluate our method; instead, we elaborated a representative subset of the query log. To do that, we first obtained all those queries which, on average, were issued at least once a day during the three months period. This way, we got about 22 thousand different queries. Secondly, we extracted those queries for which in more than half of the interactions the user did not go beyond the first page of results (i.e. presumably "easy" queries). And, finally, we obtained those queries for which in more than half of the interactions the user went indeed beyond the first page of results (i.e. presumably "difficult" queries). Thus, the final subset contained more than 28 thousand unique queries amounting for more than 9 million interactions, i.e. about a quarter of the whole query log.

Once the queries were selected we needed to resolve for each user the sequence of actions (i.e. clicks) s/he actually had performed (remember the session definition provided in a previous section). To do that, we applied a simple temporal threshold: if two actions were less than 30 minutes apart they would belong to the same session and otherwise to different sessions. Figures 1 to 3 illustrate this process from the raw query log data to the final query sessions.

```
 285103   ants   2006-04-01 19:45:23    1   http://www.dna.affrc.go.jp
 285103   ants   2006-04-01 19:45:23    3   http://www.uky.edu
 285103   ants   2006-04-01 19:50:59   13   http://ohioline.osu.edu
 285103   ants   2006-04-01 19:50:59   14   http://ohioline.osu.edu
 285103   ants   2006-04-11 21:44:45    7   http://ohioline.osu.edu

 889138   ants   2006-03-05 13:22:31    4   http://www.ants.com
 889138   ants   2006-03-05 13:22:31    8   http://home.att.net
 889138   ants   2006-03-05 13:26:14   11   http://www.infowest.com
 889138   ants   2006-03-05 13:26:14   19   http://www.greensmiths.com

3519380   ants   2006-03-30 17:14:14    0
3519380   ants   2006-03-30 17:15:53    1   http://ant.edb.miyakyo-u.ac.jp
3519380   ants   2006-03-30 17:15:53    3   http://www.uky.edu
3519380   ants   2006-03-30 17:15:53   10   http://en.wikipedia.org
3519380   ants   2006-03-30 17:27:46    0
3519380   ants   2006-04-01 13:55:03    2   http://www.lingolex.com
3519380   ants   2006-04-01 13:55:03    3   http://www.uky.edu
3519380   ants   2006-04-01 14:20:53    0
```

**Figure 1.** Fragment from the AOL query log corresponding to the query 'ants'. From left to right: user ID, query string, time stamp, position of the clicked URL in the result page, and clicked URL. A zero rank means the user just issued the query but did not click any result.

```
 285103    ants    2006-04-01 19:45:23    1     http://www.dna.affrc.go.jp
 285103    ants    2006-04-01 19:45:23    3     http://www.uky.edu
 285103    ants    2006-04-01 19:50:59    13    http://ohioline.osu.edu
 285103    ants    2006-04-01 19:50:59    14    http://ohioline.osu.edu
---------------------------------------------------------------------------
 285103    ants    2006-04-11 21:44:45    7     http://ohioline.osu.edu

 889138    ants    2006-03-05 13:22:31    4     http://www.ants.com
 889138    ants    2006-03-05 13:22:31    8     http://home.att.net
 889138    ants    2006-03-05 13:26:14    11    http://www.infowest.com
 889138    ants    2006-03-05 13:26:14    19    http://www.greensmiths.com

 3519380   ants    2006-03-30 17:14:14    0
 3519380   ants    2006-03-30 17:15:53    1     http://ant.edb.miyakyo-u.ac.jp
 3519380   ants    2006-03-30 17:15:53    3     http://www.uky.edu
 3519380   ants    2006-03-30 17:15:53    10    http://en.wikipedia.org
 3519380   ants    2006-03-30 17:27:46    0
---------------------------------------------------------------------------
 3519380   ants    2006-04-01 13:55:03    2     http://www.lingolex.com
 3519380   ants    2006-04-01 13:55:03    3     http://www.uky.edu
 3519380   ants    2006-04-01 14:20:53    0
```

**Figure 2.** The interactions from Figure 1 sessionized by using a 30 minute threshold.

```
 3519380   ants    2006-03-30 17:14:14    1     http://ant.edb.miyakyo-u.ac.jp
                                          3     http://www.uky.edu
                                          10    http://en.wikipedia.org
---------------------------------------------------------------------------
 3519380   ants    2006-04-01 13:55:03    2     http://www.lingolex.com
                                          3     http://www.uky.edu
```

**Figure 3.** Two different interactions from the same user, they include the sequence of clicks and their position within the original page of results.

### 4.1.2. Search engine evaluation

Before describing the way in which we evaluated our method against the AOL dataset, some background is needed on the performance measures usually reported for search engines.

Perhaps the simplest information retrieval performance measure is the so-called "precision" ($P$). It is simply the fraction of the provided results which are relevant. In spite of its simplicity, there are subtle nuances if we try to apply such a measure to search engine evaluation. For instance, let's suppose the following sequence of user actions: $d_1^*, d_2, d_3^*, d_4, d_5^*, d_6, d_7, d_8, d_9, d_{10}$. Even assuming the aforementioned *"Click > Skip above"* strategy we cannot resolve if documents from 6 to 10 were or not relevant for the user.

This problem can be addressed by using condensed lists which are defined by Sakai [86] as ranked lists of documents with the unjudged ones removed. In the previous example, the condensed list would be $d_1^*, d_2, d_3^*, d_4, d_5^*$ and, thus, the precision for such condensed list ($P'$) could be computed: 0.6.

It is worth noting that the work by Sakai is related to other works aiming to devise new performance measures to deal with such incomplete relevance judgments (e.g. bpref [16], or rank-biased precision [66]). A discussion on the difference between those measures is needless for this work. Suffice it to say that we decided to employ the so-called *nDCG'*[10] for two good reasons: (1) according to Sakai *nDCG'* is preferable to most of the other performance measures (including [16, 66]) because it is simpler and more robust to incomplete data; and (2) most of the recent work on Web information retrieval tend to use that measure.

---
[10] Normalized discounted cumulative gain for condensed lists.

We will now provide a brief review of *nDCG* and *DCG* in which is based [45]. The underlying idea is quite simple: as a user goes down examining a list of results his gain increases depending on the individual gain of each document (if documents can only be relevant or irrelevant the gain would be binary). This can be expressed by the following equation, where $G_i$ is the gain of the document at position *i*, and $CG_p$ is the cumulative gain up to position *p*:

$$CG_p = \sum_{i=1}^{p} G_i$$

This, however, does not entirely reflect the users' expectations, that is, they prefer relevant documents to appear sooner than later and, thus, results appearing in the last positions should provide less gain than other appearing earlier. This can be achieved this way:

$$DCG_p = CG_{b-1} + \sum_{i=b}^{p} \frac{G_i}{\log_b i}$$

The value of parameter *b* is usually assumed to be 2, thus:

$$DCG_p = CG_1 + \sum_{i=2}^{p} \frac{G_i}{\log_2 i}$$

Finally, the measure must be normalized in order to obtain performance measures in the range [0, 1] where 1 would mean ideal performance. In fact, the *nDCG* is obtained by dividing *DCG* by *IDCG* which is the *DCG* value for an ideal ranking of the results. Figure 4 shows the steps to compute $nDCG_5$ for the condensed list $d_1^*$, $d_2$, $d_3^*$, $d_4$, $d_5^*$ assuming binary relevance.

$$R = \{d_1^*, d_2, d_3^*, d_4, d_5^*\} \qquad\qquad G = \{1,0,1,0,1\}$$

$$DCG_5 = 1 + \frac{0}{\log_2 2} + \frac{1}{\log_2 3} + \frac{0}{\log_2 4} + \frac{1}{\log_2 5} = 1 + 0 + 0.631 + 0 + 0.431 = 2.062$$

$$IR = \{d_1^*, d_3^*, d_5^*, d_2, d_4\} \qquad\qquad IG = \{1,1,1,0,0\}$$

$$IDCG_5 = 1 + \frac{1}{\log_2 2} + \frac{1}{\log_2 3} + \frac{0}{\log_2 4} + \frac{0}{\log_2 5} = 1 + 1 + 0.631 + 0 + 0 = 2.631$$

$$nDCG_5 = \frac{DCG_5}{IDCG_5} = \frac{2.062}{2.631} = 0.784$$

**Figure 4.** Steps to compute $nDCG_5$ for a given condensed list.

So, in short, we can compute *nDCG* values at different positions for every query from the dataset by simply assuming that clicked results are relevant and skipped results irrelevant. In fact, by doing this we would be evaluating the performance of the search engine powering AOL which, by the way, is *Google*.

### 4.1.3. Producing synthetic usage data by means of Monte Carlo simulation

At this moment we have described the dataset obtained from the AOL query log and the way in which it is possible to use such click-through data to evaluate the performance of a search engine. However, we have not explained how we planned to exploit such data to evaluate the performance of our technique.

As it was previously described, our method requires queries, clicks and temporal data in order to evolve the pheromone weights for each query-document pair. The AOL dataset contains all that information and, thus, it can be easily used to "train" our system. Besides, given that it contains data for three months it can be divided into training and test sets with each of them still containing plenty of data.

We argued that our method can work in real-time but we feel that, for an offline experiment, it would be far more comprehensive to use one portion to simulate user behavior and the rest for evaluation than performing an accurate day-by-day simulation. Thus, two different partitions were prepared: in the first one, data from March was used for training and April and May for testing; in the second one, March and April were used for training and May for testing.

In addition to different training data we were interested in evaluating different parameters such as the decay time ($\delta$ in the model) and the number of results to recommend. We tested the method with two different $\delta$ values (one day and one week), and recommending 1 and 3 documents. This, combined with the three different flavors, and the two different data partitions gave to 24 different runs.

Each of those runs was a Monte Carlo simulation because, as we have said, the result recommendation process is stochastic. Thus, for each user interaction to simulate, 10 possible outcomes were produced. Then, each of those outcomes was compared with the actual user interaction to obtain the alleged clicks that given user would have produced if s/he had found that different ordering of the results. The basic assumption under this comparison was simple: if a given user, during one specific session, had clicked a document in a position $p$ as a result for a query, we can safely assume that the same user in the same session would have clicked the same document in any position above $p$.

Figure 5 will help to illustrate this. The user **307464** issued the query **'ants'** at a given time and day; in that session, that user clicked two results appearing in positions 1 and 10, respectively. In the middle of the Figure we show the Monte Carlo simulation for that particular session; as it can be seen, depending on the iteration different URLs are recommended. Let's now take the first iteration, where the URL **http://ohioline.osu.edu** is suggested. As we have previously mentioned our method "injects" recommendations as top results; in this particular case it means that this recommendation would push **http://www.lingolex.com** down from the first to the second place. Besides, given that the recommendation appears in the original click list for the session in a later position, we know for sure that the user would click on it. Thus, while in the actual session the user had clicked results 1 and 10 (**lingolex** and **ohioline**, respectively), in the swarm-based search engine he would have clicked results 1 and 2 (this time, **ohioline** and **lingolex**). Hence, for this iteration our method would outperform the actual search engine. Finally, for every session the performance of the different iterations is averaged.

**Real user session**

```
307464    ants    2006-05-20 21:48:49    1       http://www.lingolex.com
                                         10      http://ohioline.osu.edu
```

**Monte Carlo simulation for that session**

```
307464    ants    2006-05-20 21:48:49    iter-0    http://ohioline.osu.edu
307464    ants    2006-05-20 21:48:49    iter-1    http://www.lingolex.com
307464    ants    2006-05-20 21:48:49    iter-2    http://www.lingolex.com
307464    ants    2006-05-20 21:48:49    iter-3    http://www.lingolex.com
307464    ants    2006-05-20 21:48:49    iter-4    http://ohioline.osu.edu
307464    ants    2006-05-20 21:48:49    iter-5    http://www.lingolex.com
307464    ants    2006-05-20 21:48:49    iter-6    http://ohioline.osu.edu
307464    ants    2006-05-20 21:48:49    iter-7    http://www.lingolex.com
307464    ants    2006-05-20 21:48:49    iter-8    http://www.lingolex.com
307464    ants    2006-05-20 21:48:49    iter-9    http://ohioline.osu.edu
```

**Original user clicks vs. alleged clicks under recommendation**

```
307464    ants    2006-05-20 21:48:49    iter-0    1,10    1,2
307464    ants    2006-05-20 21:48:49    iter-1    1,10    1,10
307464    ants    2006-05-20 21:48:49    iter-2    1,10    1,10
307464    ants    2006-05-20 21:48:49    iter-3    1,10    1,10
307464    ants    2006-05-20 21:48:49    iter-4    1,10    1,2
307464    ants    2006-05-20 21:48:49    iter-5    1,10    1,10
307464    ants    2006-05-20 21:48:49    iter-6    1,10    1,2
307464    ants    2006-05-20 21:48:49    iter-7    1,10    1,10
307464    ants    2006-05-20 21:48:49    iter-8    1,10    1,10
307464    ants    2006-05-20 21:48:49    iter-9    1,10    1,2
```

**Figure 5.** Monte Carlo simulation of the recommendation process and generation of the alleged user clicks from the original click-through data.

To sum up, the goal of the offline experiment was to compare the performance of the different flavors of our technique under different parameterizations with the performance of a major search engine. We have outlined a method to carry out such a comparison by using a huge query log. Given that our technique relies on user interaction to discover the most promising results, the query log has to be divided into training and test sets. With regards to the testing, we have described (1) the way in which Monte Carlo simulation is applied to produce a number of outcomes for every real session in the test set; and (2) how such outcomes can be combined with the original data to produce an alleged click sequence that such user would exhibit provided s/he had received the given recommendation. In a later section the results obtained by our method within this evaluation framework are provided.

### 4.2. Online evaluation

In addition to the offline evaluation we considered that an online one by means of a controlled experiment was also needed for several reasons. First, as we have previously exposed, the real-time features of the method were not addressed in the offline experiment. Second, given the strict assumptions to generate the synthetic data, it was not possible to determine the impact our method could have on those users who just focus in the first results. And, third, no matter which performance increase the offline evaluation could reveal, if it did not have a measurable impact on real users it would be useless [41, 95].

Thus, the controlled experiment was designed to test our hypothesis that users of a swarm-based adaptable search engine would gradually require less time to find information, provided other users have sought for it previously.

Therefore, we deployed two instrumented search engines with *Google*'s look-and-feel (see Figure 6) although they were based on *Yahoo! Search BOSS*[11]. In addition to that, the computers in the lab used for the experiment were configured to resolve `google.com` to our server's IP address. This way we could be certain of the users were unable to detect the ploy. One of the search engines simply recorded the users' actions. The second one, in addition to that, implemented our technique in the *Naïve* version[12] and recommended up to three results (provided they were available) for each query. The recommended results were indistinguishable from the other results in the page and, thus, the users did not have any incentive to click on those results other than their appeal to solve the task in course.

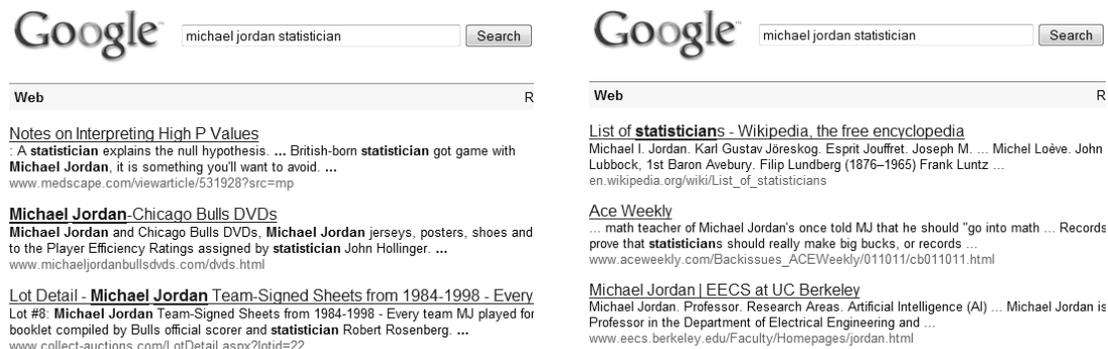

**Figure 6.** On the left the instrumented search engine, on the right the swarm-based search engine once the controlled experiment had concluded. In this case the task consists of finding the homepage of the statistician Michael Jordan; the first result on the right pane is highly relevant to the task, and the third one is a possible answer.

The experiment involved 20 subjects (faculty and students from our Department) who were persuaded to believe it dealt with human-computer interaction during Web searching. The subjects were randomly assigned to the control and experimental groups and each subject took part in the experiment alone (except for the researcher conducting the experiment).

Each participant had to solve 12 different tasks (see Table 1), 9 of them were taken from the study by Joachims *et al.* [48][13], the other three were proposed by these authors. 4 of the tasks were considered trivial, the rest of the tasks were more difficult although the degree of difficulty varied from one task to another. Prior to conducting the experiment, the tasks were randomly ordered and, then, all the subjects solved the tasks in that order; besides, they could only read a task description after answering the previous one. Participants were instructed to solve the tasks employing (what they thought it was) *Google* as they usually did, and answer the questions verbally to the researcher. In case their tentative answer was incorrect they were told so, and they had to try again. The users had no time limit to solve the tasks and were informed of this.

---

[11] *Yahoo! Search BOSS (Build your Own Search Service)* is a web service platform that allow developers to build applications on top of the *Yahoo!* platform; for instance, it offers programmatically access to their search engine. `http://developer.yahoo.com/search/boss/`

[12] For he controlled experiment we introduced a minimal change to the system: instead of using query-documents pairs –e.g. `(michael jordan statistician, URL)`–, we generate a pair for every combination of terms within the query and the URL –e.g. `(michael, URL)`, `(michael jordan, URL)`, `(jordan statistician, URL)`, etc. By doing this we were addressing the potential vocabulary mismatch between different participants.

[13] Joachims et al. [48] provided 10 tasks; we removed one that, by 2009, was outdated: "*With the heavy coverage of the democratic presidential primaries, you are excited to cast your vote for a candidate. When are democratic presidential primaries in New York?*"

To sum up, participants from the control group were using *Yahoo!* search engine disguised as *Google*, while the experimental group was working with the swarm-based search engine. All of them were solving the same tasks on their own but, unlike the users in the control group which were actually alone, the subjects in the experimental group were unknowingly taking advantage of the actions of previous participants. Thus, the expected output for the experiment is clear: we should detect changes in the time required to solve tasks in the experimental group, and such changes should relate to their position within the group (i.e. the later the subject use the search engine the easier the tasks should be). The results obtained in this experiment are provided in the following section.

| greyhound *(task 1)* | cornell *(task 2)* |
|---|---|
| Find the page displaying the route map for Greyhound buses. | The founder of Cornell University used to live close to campus –near University and Stewart Avenue. Does anybody live in this house now? If so, who? |
| *Joachims et al. [48].* | *Joachims et al.[48]. Slightly rephrased.* |
| A priori trivial. | A priori difficult. |
| baeza *(task 3)* | time-machine *(task 4)* |
| Find the homepage of Ricardo Baeza, not the director of Yahoo! Research Barcelona but another fellow countryman of him. | Which actor starred as the main character in the original Time Machine Movie? |
| *By the authors.* | *Joachims et al. [48].* |
| A priori difficult. | A priori difficult. |
| ny-mountains *(task 5)* | purple-cow *(task 6)* |
| Which are the tallest mountains in New York? | How can you follow the international bestselling author of Purple Cow on Twitter? |
| *Joachims et al. [48]. Slightly rephrased.* | *By the authors.* |
| A priori difficult. | A priori difficult. |
| 1000-acres *(task 7)* | antibiotic *(task 8)* |
| Find the homepage of the 1000 Acres Dude Ranch. | What is the name of the researcher who discovered the first modern antibiotic? |
| *Joachims et al. [48].* | *Joachims et al. [48].* |
| A priori trivial. | A priori difficult. |
| michael-jordan *(task 9)* | emeril *(task 10)* |
| Find the homepage of Michael Jordan, the statistician. | Find the homepage of Emeril, the chef who has a television cooking program. |
| *Joachims et al. [48].* | *Joachims et al. [48].* |
| A priori difficult. | A priori trivial. |

strangelove *(task 11)*

This picture of Homer Simpson is a reference to a famous scene from a classic movie.
- What is the title of that episode?
- And the title of the movie?
- What's the name of the character in the original movie?
- And the name of the actor who played it?

*By the authors.*
A priori difficult.

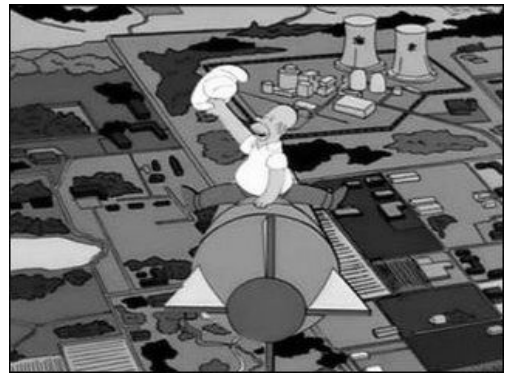

cmu *(task 12)*

Find the homepage for graduate housing at Carnegie Mellon University.

*Joachims et al. [48].*

A priori trivial.

**Table 1.** The 12 tasks used for the controlled experiment. We include the code-name for the tasks, task description as was provided to the subjects, source of the task, and the a priori difficulty of the task according to these authors. Needless to say, participants in the experiment were only aware of the task description.

# 5. Results

This study was driven by three research questions. The first one deals with the feasibility of applying a swarm-based approach to search engines in order to exploit users' behavior to achieve better results. This question has been partially addressed in section 3.2 where a model for a swarm-based search engine is proposed; however, to completely resolve the feasibility of such a technique we must answer the second and third research questions. The second one, dealing with the performance increase achieved with this method is addressed in the following subsection regarding the results of the offline experiment. The third one, concerning the impact on real users' experience is later addressed in the subsection describing the results of the controlled experiment.

## 5.1. Offline experiment

As it was exposed in the Research design, we devised an experimental framework to test our method against a representative subset of interactions extracted from the AOL query log. We described (1) the way in which such subset was divided into training and test sets; (2) how the training set was employed to simulate real users' interactions with our swarm-based search engine; (3) the way in which recommendations were produced by our method; and (4) a manner to produce synthetic usage data by combining such recommendations and the original users' click-through data.

In addition to that, we briefly discussed the performance measures that can be applied to evaluate search engines and the reasons to choose one specifically, namely the normalized discounted cumulative gain (*nDCG*).

Finally, we noted that 24 different Monte Carlo simulations were run by using different flavors of the technique, different sizes for the training and test sets, and different parameterizations. For the sake of brevity we do not provide results for all those experiments, but only for the parameterization achieving best results, i.e. training on data from March to April, applying a decay factor of 24 hours, and recommending just 1 result. Further details and analysis of both the different settings and the results achieved with them is later provided in the Discussion section.

At this point, we must make a note regarding the nature of the queries. There is broad consensus within the field of Web Information Retrieval that, when interacting with a search engine, users' reveal one of three possible intents [15]: (1) *"navigational"* (the user wants to reach a particular website), (2) *"informational"* (the user wants to find a piece of information on the Web), and (3) *"transactional"* (the user wants to perform a web-mediated task). An example of a navigational query would be `greyhound route map` (which could be used to solve the first task from the controlled experiment). An informational query could be, for instance, `who discovered first antibiotic` (which could appear during the eighth task). Finally, `cheap broadway tickets` could exemplify transactional queries[14].

By definition, navigational queries should have only one possible correct result and, thus, in the great majority of cases the search engine's first result is the correct "answer" to such kind of queries. Thus, little to no performance increase should be expected to be achieved by our technique regarding navigational queries. On the other hand, transactional and, even more, informational queries are much vaguer and, hence, more difficult to solve for the search engine. Consequently, we expected most of the performance increase to be found in this kind of queries.

---
[14] None of the tasks in the controlled experiment was supposed to require such kind of queries because the user was expected to give answers, not to perform actions.

Thus, in order to determine if our method had or not any impact on non-navigational queries some method to automatically detect them was needed. There exist several techniques to detect query intent [13]; nevertheless, one of the simplest is the one proposed by Jansen *et al.* [42]. It consists of a number of easily implementable rules which can be summarized as follows: navigational queries contain names of companies, businesses, organizations, or people; they contain domains suffixes; or they have less than three terms.

Some of these rules require external information and, to that end, several lists of pertinent terms were obtained from Freebase[15] by means of MQL [32] queries (see Figures 7 and 8). That way, we obtained lists of companies, organizations, websites, and people names and surnames. Figure 9 shows some of such terms. Then, we applied the rule-base classifier to divide the results in two sets: navigational and non-navigational.

```
{
  "cursor":true,
  "query":[
    {
      "key":[],
      "name":[],
      "type":"/business/company"
    }
  ]
}
```

**Figure 7.** A MQL query to retrieve the name and keys of all the companies available in Freebase.

```
{
  "key" : [
    "848",
    "Audi",
    "Audi_AG",
    "audi",
    "Audi_Aktien-Gesellschaft",
    "Audi_Sport"
  ],
  "name" : [
    "Audi"
  ],
  "type" : "/business/company"
}
```

**Figure 8.** One "record" obtained with the previous query.

```
alsa bus company       craigslist
cajastur               digg
microsoft corporation  john
uc los angeles         william
uk labour party        james
unicef                 moore
blogger                jackson
```

**Figure 9.** Some of the terms employed to implement the technique by Jansen *et al.* They are companies, organizations, websites, and people names.

---

[15] **http://www.freebase.com/**

There is one last thing to mention with regards to the averaging of results. In order to provide a final figure summarizing the performance of the system the results must be averaged which, according to Lewis [53], can be done in two ways: micro and macro. When macro-averaging, each individual experiment is considered separately and the average performance is computed from the individual performance figures obtained within each experiment. When micro-averaging results, the data from different experiments is considered to belong to one unique larger experiment.

Thus, with regards to the evaluation reported in this section, micro-averaging consists in computing the *nDCG* for every single interaction in the dataset and then averaging all of them. This way, most active users and most frequent queries dominate the results; in other words, micro-averaged results are equivalent to a weighted-average both by users and queries.

On the other hand, macro-averaging consists in computing the average measures from those obtained in every individual experiment; in this sense we have grouped data by query and by user to offer two different points of view of the system performance.

So, in short, the following tables show the performance achieved by our technique in its three different flavors under the best parameterization. We are providing both micro- and macro-averaged results for three different datasets: the whole dataset (i.e. without separating navigational from non-navigational queries), navigational (i.e. easy) queries, and non-navigational (i.e. hard) queries.

|  | Baseline search engine | Naïve | Δ(%) | Ranking bias | Δ(%) | Elaborate | Δ(%) |
|---|---|---|---|---|---|---|---|
| *nDCG@1* | 0.411938 | 0.364222 | -11.58% | 0.346156 | -15.97% | 0.385407 | -6.44% |
| *nDCG@3* | 0.457445 | **0.464462** | **1.53%** | **0.464985** | **1.65%** | **0.463880** | **1.41%** |
| *nDCG@10* | 0.477946 | **0.482568** | **0.97%** | **0.482830** | **1.02%** | **0.482277** | **0.91%** |

**Table 2.** Micro-averaged results (whole query subset)

|  | Baseline search engine | Naïve | Δ(%) | Ranking bias | Δ(%) | Elaborate | Δ(%) |
|---|---|---|---|---|---|---|---|
| *nDCG@1* | 0.475351 | 0.427299 | -10.11% | 0.407275 | -14.32% | 0.445977 | -6.18% |
| *nDCG@3* | 0.517911 | **0.521617** | **0.72%** | **0.521958** | **0.78%** | **0.521263** | **0.65%** |
| *nDCG@10* | 0.531478 | **0.533950** | **0.47%** | **0.534117** | **0.50%** | **0.533760** | **0.43%** |

**Table 3.** Micro-averaged results (navigational queries)

|  | Baseline search engine | Naïve | Δ(%) | Ranking bias | Δ(%) | Elaborate | Δ(%) |
|---|---|---|---|---|---|---|---|
| *nDCG@1* | 0.280837 | 0.233815 | -16.74% | 0.219796 | -21.74% | 0.260181 | -7.36% |
| *nDCG@3* | 0.332434 | **0.346297** | **4.17%** | **0.347197** | **4.44%** | **0.345244** | **3.85%** |
| *nDCG@10* | 0.367270 | **0.376338** | **2.47%** | **0.376796** | **2.59%** | **0.375839** | **2.33%** |

**Table 4.** Micro-averaged results (non-navigational queries)

|  | Baseline search engine | Naïve | Δ(%) | Ranking bias | Δ(%) | Elaborate | Δ(%) |
|---|---|---|---|---|---|---|---|
| *nDCG@1* | 0.314770 | 0.275173 | -12.58% | 0.259984 | -17.41% | 0.292558 | -7.06% |
| *nDCG@3* | 0.355584 | **0.361309** | **1.61%** | **0.361863** | **1.77%** | **0.360693** | **1.44%** |
| *nDCG@10* | 0.375868 | **0.379585** | **0.99%** | **0.379877** | **1.07%** | **0.379269** | **0.90%** |

**Table 5.** Macro-averaged results with respect to user (whole query subset)

|           | Baseline search engine | Naïve    | Δ(%)    | Ranking bias | Δ(%)    | Elaborate | Δ(%)    |
|-----------|-----------------------|----------|---------|--------------|---------|-----------|---------|
| nDCG@1    | 0.368598              | 0.329410 | -10.63% | 0.312201     | -15.30% | 0.345025  | -6.40%  |
| nDCG@3    | 0.406663              | **0.409980** | 0.82% | **0.410354** | 0.91%   | 0.409560  | 0.71%   |
| nDCG@10   | 0.421133              | **0.423294** | 0.51% | **0.423490** | 0.56%   | 0.423065  | 0.46%   |

**Table 6.** Macro-averaged results with respect to user (navigational queries)

|           | Baseline search engine | Naïve    | Δ(%)    | Ranking bias | Δ(%)    | Elaborate | Δ(%)    |
|-----------|-----------------------|----------|---------|--------------|---------|-----------|---------|
| nDCG@1    | 0.225838              | 0.179253 | -20.63% | 0.166625     | -26.22% | 0.202457  | -10.35% |
| nDCG@3    | 0.272968              | **0.283754** | 3.95% | **0.284660** | 4.28%   | 0.282617  | 3.53%   |
| nDCG@10   | 0.306437              | **0.313354** | 2.26% | **0.313829** | 2.41%   | 0.312803  | 2.08%   |

**Table 7.** Macro-averaged results with respect to user (non-navigational queries)

|           | Baseline search engine | Naïve    | Δ(%)    | Ranking bias | Δ(%)    | Elaborate | Δ(%)    |
|-----------|-----------------------|----------|---------|--------------|---------|-----------|---------|
| nDCG@1    | 0.336780              | 0.291300 | -13.50% | 0.279851     | -16.90% | 0.316347  | -6.07%  |
| nDCG@3    | 0.382211              | **0.396669** | 3.78% | **0.397536** | 4.01%   | 0.395187  | 3.39%   |
| nDCG@10   | 0.414054              | **0.423600** | 2.31% | **0.424044** | 2.41%   | 0.422800  | 2.11%   |

**Table 8.** Macro-averaged results with respect to query (whole query subset)

|           | Baseline search engine | Naïve    | Δ(%)    | Ranking bias | Δ(%)    | Elaborate | Δ(%)    |
|-----------|-----------------------|----------|---------|--------------|---------|-----------|---------|
| nDCG@1    | 0.451331              | 0.399449 | -11.50% | 0.385183     | -14.66% | 0.422403  | -6.41%  |
| nDCG@3    | 0.494363              | **0.503107** | 1.77% | **0.503628** | 1.87%   | 0.502324  | 1.61%   |
| nDCG@10   | 0.515387              | **0.521244** | 1.14% | **0.521512** | 1.19%   | 0.520819  | 1.05%   |

**Table 9.** Macro-averaged results with respect to query (navigational queries)

|           | Baseline search engine | Naïve    | Δ(%)    | Ranking bias | Δ(%)    | Elaborate | Δ(%)    |
|-----------|-----------------------|----------|---------|--------------|---------|-----------|---------|
| nDCG@1    | 0.257291              | 0.216253 | -15.95% | 0.206758     | -19.64% | 0.242752  | -5.65%  |
| nDCG@3    | 0.304387              | **0.322810** | 6.05% | **0.323916** | 6.42%   | 0.320843  | 5.41%   |
| nDCG@10   | 0.343737              | **0.355843** | 3.52% | **0.356409** | 3.69%   | 0.354783  | 3.21%   |

**Table 10.** Macro-averaged results with respect to query (non-navigational queries)

### 5.2. Controlled experiment

Once the controlled experiment was finished, we processed the data collected in the system logs to compare both the queries each group issued and the time required to complete the tasks.

According to our hypothesis, users of the swarm-based adaptable search engine would require less time to solve their tasks because they would unknowingly take advantage from the actions of previous users. Thus, by all accounts their queries should be similar to those submitted not only by other subjects within the experimental group but similar to those from users in the control group. To find those similarities we relied on the well-known cosine similarity which was computed at two different levels: using all the queries issued by the user to solve each task, and using just the first query. In addition to that, we computed not only the similarity between users within the same group but also cross-group similarities. Figures 10 and 11 show the average cosine similarity for each of the different tasks.

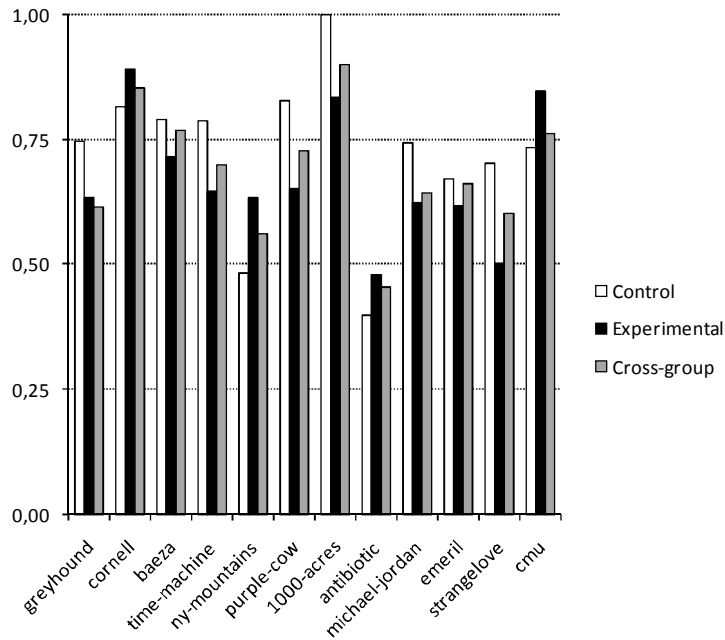

**Figure 10.** Average cosine similarity between the first queries within user sessions.

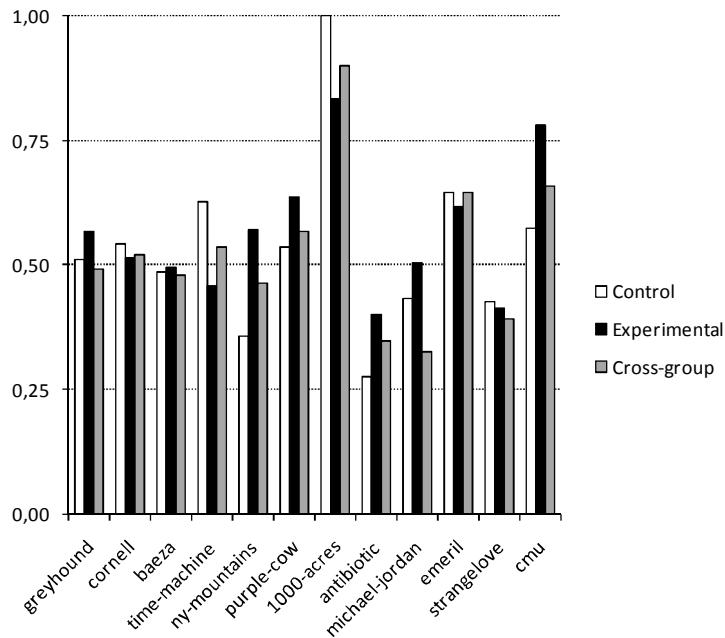

**Figure 11.** Average cosine similarity between complete user sessions.

However, the most important result that the controlled experiment was expected to reveal was the evolution of the time needed to solve the tasks as more and more users employed the swarm-based system. If our initial hypothesis was true, the data should reveal a strong negative correlation between the time needed to solve the tasks and the order in which a user from the experimental group took part in the experiment. Table 11 shows the Pearson's correlation coefficient between time required to solve the tasks, and order of participation; data is provided for both total and average time and it is separated into trivial and non-trivial tasks.

|  | Control group | | Experimental group | |
| --- | --- | --- | --- | --- |
|  | **Trivial tasks** | **Non-trivial tasks** | **Trivial tasks** | **Non-trivial tasks** |
| **Total time** | 0.093730 | 0.379485 | *-0.612676* | **-0.713765** |
| **Average time** | 0.094308 | 0.380008 | *-0.611763* | **-0.714156** |

**Table 11.** Pearson's *r* correlation coefficient between the time required to complete the tasks and the position of the subjects within their reference group. Grey cells are those statistically significant (italics at the 10% level and bold at the 5% level).

Finally, box-and-whisker plots are provided (see Figures 12 and 13) to allow comparison between the time required by users to solve trivial and non-trivial tasks, and also to compare the differences between the control and experimental groups. Further discussion on these results is provided in the following section.

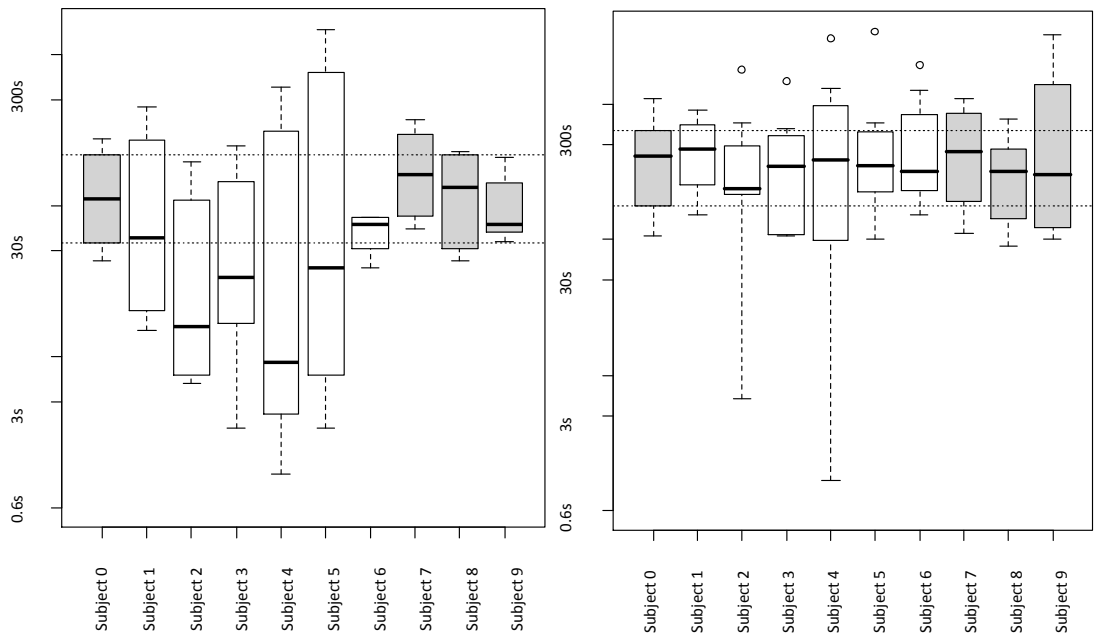

**Figure 12.** Box-and-whisker plots for the control group, time is shown in log scale. The left pane shows the data corresponding to the trivial tasks while the right pane corresponds to non-trivial tasks.

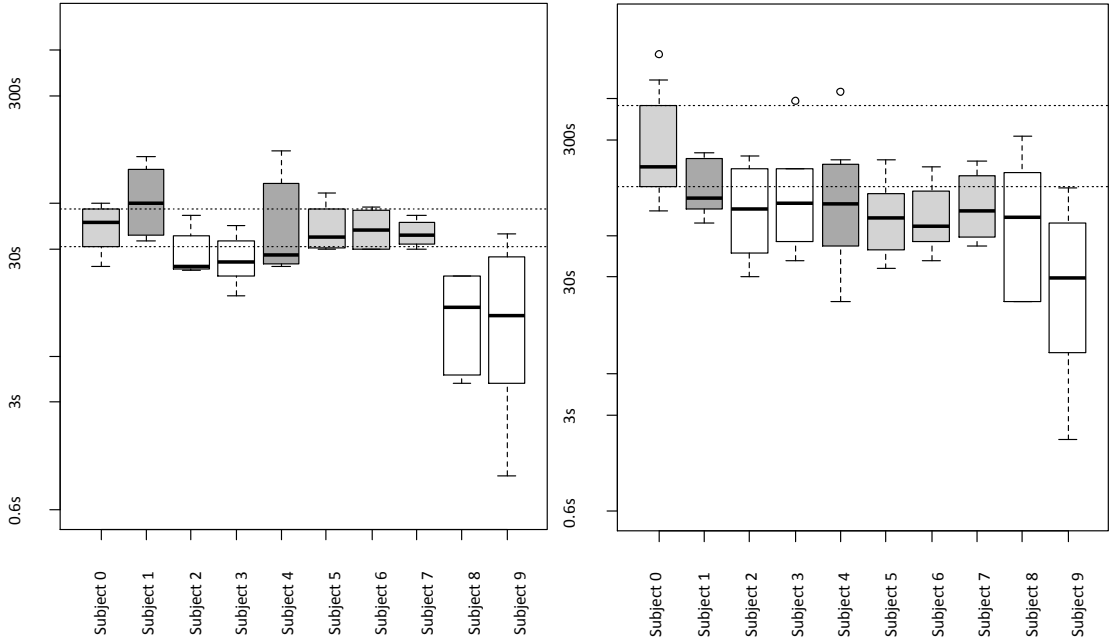

**Figure 13.** Box-and-whisker plots for the experimental group, time is shown in log scale. The left pane shows the data corresponding to the trivial tasks while the right pane corresponds to non-trivial tasks.

## 6. Discussion of results

At this point we should return to the research questions driving this study: (1) Is it feasible to implement a swarm-based real-time adaptable search engine? (2) For which kind of queries, if any, does the search engine's performance increase? And, (3) does the user experience improve in any measurable way? As we have already said, once a model for a swarm-based search engine was proposed the answer to the first question would lie on the other two and those will be properly addressed in the following two subsections.

### 6.1. Offlline experiment

In the Results section we noted that only results from the best parameterization were provided; the results for the other parameterizations were quite mixed and they greatly depended on the parameters coming into play: namely, the size of the training set, the decay factor, and the number of results to recommend.

It is barely surprising that the performance is superior when using the largest training set (i.e. data from March and April); this seems expectable from a swarm-based approach: the more usage behavior is collected the better the performance of the system. What seemed slightly unexpected was the differences between different decay factors ($\delta$). We had expected that using $\delta$=604800 (i.e. a week) would provide better performance because weekly cycles amount for a great number of queries [58]; however, using $\delta$=86400 (i.e. a day) was slightly better. Regarding the number of results to recommend, from the offline experiment one should expect poor results by providing more than one recommendation. However, we feel that this may be expectable given the way in which the offline evaluation had to be done; in fact, for the online controlled experiment we configured the system to recommend up to three results.

So, in short, our technique (independently of the flavor) achieved best results when learning from two months, having a decay factor of one day and recommending just one document per query. Therefore, we will just discuss the results obtained under those conditions.

As it can be seen in Tables 2 to 10, we are reporting *nDCG* values at three different positions: at 1, 3, and 10 results (i.e. *nDCG@1*, *nDCG@3* and *nDCG@10*). As it can be seen, the performance of the method drops when reporting *nDCG@1*, reveals the highest gains for *nDCG@3* and still exhibits better performance than the baseline search engine for *nDCG@10*. Such results are entirely expected and consistent with an acceptable performance of the proposed method.

Regarding the seemingly poor performance for *nDCG@1*: Let's remember that, in the offline experiment, we just can assume that the users would have clicked the same results they had originally clicked as long as they were provided in a higher ranking. It is impossible to know if they would have clicked a document from a lower rank that they had not previously examined had it appeared as a top result. Thus, the only way any method could increase baseline performance at the first result when evaluating in this manner would be by consistently recommending the actual document the user was going to click at that given session. Needless to say, such a thing would be fantastic but it is mostly unachievable.

Thus, we should focus in the performance reported by both *nDCG@3* and *nDCG@10*. As it can be seen, our method outperforms the baseline search engine consistently. Also, it is not unexpected that the greatest boost at performance is achieved when just evaluating the first 3 results rather than the whole 10 results in the page, because our system was just providing one recommendation.

With regards to the kind of queries with a larger performance gain, the results reveal that, as expected, the system has little to no impact on navigational queries while it boosts the performance for non-navigational ones. Regarding the differences between the different flavors of the technique, they are mostly negligible although the *Ranking-bias* version consistently outperforms the others.

Finally, we must address the question about the significance of these results. A criterion commonly applied in Information Retrieval is the one proposed by Spärck-Jones [91]: performance differences lesser than 5% should be disregarded, those in the 5-10% interval are "noticeable", and "material" only those greater than 10%.

According to such a criterion the only significant results would be those shown in Table 10 (non-navigational data macro-averaged with respect to query); this would mean that the system is actually having an impact on the performance for navigational queries. However, both micro-averaged results for non-navigational queries (Table 4) and macro-averaged results with respect to user for non-navigational queries (Table 7) are close to that threshold (4.44% and 4.28% respectively for the Ranking-bias version).

Therefore, we feel it can be safely concluded that, in effect, a swarm-based approach can be applied to search engines with a significant impact on the performance for non-navigational queries which are, indeed, the hardest to solve at this moment.

### 6.2. Controlled experiment

There is still one remaining question: does the proposed technique have any measurable impact on real users' search experience? The controlled experiment aimed to answer that was described in the Research design section. It involved two groups of participants who had to solve several tasks using what they thought was *Google*. The control group used a major search engine (*Yahoo* in a *Google* disguise) while the experimental group used a search engine empowered by our technique. Both search engines collected the users' interactions which were later processed to obtain results which were provided in the previous section.

As we have previously explained, we were interested in two main aspects of the participants' behavior: namely, the queries they issued, and the time they required to solve the tasks. The main underlying assumption of the proposed technique is that users can profit from previous actions from other individuals unknowingly and, thus, their behavior, mostly revealed in their queries, should suffer minor changes or, even, not change at all. Therefore, in the previous section we have shown the average cosine similarity between users within each reference group and between users from both groups. Figures 10 and 11 reveal that differences between both groups for each task are mostly negligible –in particular if we just observe the first query (all the users issued at least one query for each tasks)– and, hence, it seems reasonable to conclude that participants issued very similar queries without regards to the search engine they were using.

Regarding the time required to solve the tasks, Table 11 is pretty self-explanatory. There is no significant correlation between the order in which participants from the control group took part in the experiment and the time (neither total nor average) they required to solve both kinds of tasks. However, there exists a significant negative correlation between order and the time to complete the tasks in the experimental group; i.e. the later an individual solves a task with the swarm-based search engine the lesser time s/he needs to solve it, because several other people solved the same task before.

Of course, one could play the devil's advocate and argue that the users in the experimental group behave in such a way just by chance; however, at the 5% level the negative correlation between order and time is only significant for non-trivial tasks, not for the trivial ones. Certainly, there exist a negative correlation for trivial tasks at the 10% level but we feel that this is a by-product of the experiment: when confronted with the a priori non-trivial tasks participants from the experimental group were expecting precisely that, hard-to-solve problems; then, when realizing that the task were indeed pretty easy they faced a priori trivial tasks (which, remember, appeared intermingled within the task list) in a much more confident (and speedier) manner[16].

So, in short, it seems that there exists support for the hypothesis that the proposed technique has indeed a measurable impact on users' experience: they require less time to solve difficult informational tasks by means of a search engine.

In addition to that, the box-and-whisker plots provide also an interesting lecture on the controlled experiment. Let's first focus on Figure 12 which shows the time required by the participants in the control group to solve the trivial and non-trivial tasks. We have shadowed the first subject from the group and those other participants with a similar behavior regarding the trivial tasks[17]. If we compare their behavior regarding non-trivial tasks we can see that they are rather similar. Another aspect worth to notice in the box plot for non-trivial tasks is that half of the subjects show outliers.

---

[16] An anecdote from the controlled experiment: the last user in the experimental group ask the author conducting the experiment what he was expecting to find by asking people to answer questions that could be solved by just writing some keywords in *Google* and clicking the first result.

[17] The reader has probably noticed the important differences among users in the control group to solve the trivial tasks. This was an unexpected outcome but, as it was the case with participants from the experimental group, we also attribute this to the fact that trivial and non-trivial tasks were mixed. In the case of the control group, after facing some of the non-trivial tasks the participants likely thought there was something fishy with the trivial ones and, hence, they devoted much more time than what it was really needed.

Now, let's examine the box plots for the experimental group (Figure 13). As we have already mentioned, the time required to solve trivial tasks is rather short (at least when compared with participants from the control group). Again, we have shadowed the first subject from the group and those other individuals with a similar behavior for trivial tasks. In this case, we have also shadowed (in a darker shade) two subjects who devoted far more time to those tasks than the first subject.

If we compare the times required to solve the non-trivial tasks we observe that (1) the three individuals who were expected to behave similarly to the first subject required much less time to solve the tasks; and (2) even those two participants who were expected to devote more time required indeed less. In fact, if we pay attention to the median time we will see it steadily decreases as participants take part in the experiment. Besides, only three subjects show outliers; one of them is the first participant and the outlier tasks for the other two individuals were undeniably shorter than the maximum non-outlier time devoted by the first subject.

So, to sum up, the box plots support the conclusions we have previously derived from Table 11: users in the experimental group required less time to solve the tasks as they took part in the experiment. Such a trend is clear for non-trivial (i.e. informational) tasks and even trivial tasks seem to experience a gain[18].

## 7. Implications and conclusion

This study has addressed the broad problem of information overload focusing on one of its most pervasive manifestations nowadays: Web searching. As we have noted, this area has still much room for improvement and we have surveyed some of the main lines of research addressing the matter. The approach we have described in this work relates to previous work regarding personalization, information foraging and recommendation. Yet, we faced the problem from an unexplored angle: a swarm-based approach to search engine adaptation.

Thus, this work has several important contributions. First, it provides a broad review of the information overload problem, in general, and a more thorough review of the methods and techniques to address it regarding Web search, in particular. Second, it relates theories on users' information seeking behavior with swarm intelligence methods, specifically ant algorithms and, in this regard, it provides a model to adapt ant colony optimization methods to swarms of users employing a search engine. Third, it states two key aspects on the application of such a technique to search engines: namely, it must increase the search engine performance regarding informational queries, and it must show a measurable impact in user experience. Fourth, both offline and online evaluation methods are described and results from such experiments are provided and discussed; showing that, indeed, ant algorithms are a feasible technique to build search engines able to adapt in real-time to the behavior of their users while improving both performance and the users' search experience (at least with regards to the time required to solve complex tasks).

---

[18] The smaller gain attained with trivial tasks is totally expected: all of them were pure navigational tasks and, thus, the time required to read the tasks, enter the query, read the snippet and click the result, mostly depend on the individual user capacities and will have a lower bound.

This study also has limitations. First, the prototype is prone to spamming because of its very own nature (i.e. malicious swarms of users could easily boost certain results for certain queries). Second, the session definition assumed for the described implementation (i.e. a single query with its corresponding clicks) is rather limiting, and rich semantic relations are being missed by not taking into account related successive queries. Third, the stochastic approach to recommend results is rather naïve. And, fourth, although such a technique seems well fitted to offer personalized search that line of research has remained unexplored.

Hence, further research is needed in the following lines: (1) accurate ways to automatically detect fraud usage (e.g. [64, 65, 79]); (2) exploiting query co-occurrence within topical sessions [36]; (3) adaptation of edge selection methods from ant algorithms to improve stochastic recommendation (e.g. [21, 29]); and (4) personalization, at least in a simple way by grouping users by IP address sub-ranges (as suggested by Mei and Church [62]).

## 8. Acknowledgements

The authors would like to thank the people who took part in the controlled experiment and, particularly, Miguel Fernández. This work was partially financed by grant UNOV-09-RENOV-MB-02 from the University of Oviedo.